\documentclass[aps,preprint,prl,superscriptaddress]{revtex4}

\usepackage{graphicx}
\usepackage{amsmath}
\usepackage{amssymb}
\usepackage{units}
\usepackage{color}

\usepackage{setspace}

\begin{document}

\title{Tracking primary thermalization events in graphene with photoemission at extreme timescales}

\author{I. Gierz}
\email{Isabella.Gierz@mpsd.mpg.de}
\affiliation{Max Planck Institute for the Structure and Dynamics of Matter, Center for Free Electron Laser Science, Hamburg, Germany}
\author{F. Calegari}
\affiliation{Max Planck Institute for the Structure and Dynamics of Matter, Center for Free Electron Laser Science, Hamburg, Germany}
\affiliation{Institute for Photonics and Nanotechnologies, IFN-CNR, Milano, Italy.}
\author{S. Aeschlimann}
\author{M. Ch{\'a}vez Cervantes}
\affiliation{Max Planck Institute for the Structure and Dynamics of Matter, Center for Free Electron Laser Science, Hamburg, Germany}
\author{C. Cacho}
\author{R. T. Chapman}
\author{E. Springate}
\affiliation{Central Laser Facility, STFC Rutherford Appleton Laboratory, Harwell, United Kingdom}
\author{S. Link}
\author{U. Starke}
\author{C. R. Ast}
\affiliation{Max Planck Institute for Solid State Research, Stuttgart, Germany}
\author{A. Cavalleri}
\affiliation{Max Planck Institute for the Structure and Dynamics of Matter, Center for Free Electron Laser Science, Hamburg, Germany}
\affiliation{Department of Physics, Clarendon Laboratory, University of Oxford, Oxford, United Kingdom}

\date{\today}

\begin{abstract}
{\bf Direct and inverse Auger scattering are amongst the primary processes that mediate the thermalization of hot carriers in semiconductors. These two processes involve the annihilation or generation of an electron-hole pair by exchanging energy with a third carrier, which is either accelerated or decelerated. Inverse Auger scattering is generally suppressed, as the decelerated carriers must have excess energies higher than the band gap itself. In graphene, which is gapless, inverse Auger scattering is instead predicted to be dominant at the earliest time delays \cite{Winzer2010,Winzer2012,Brida2013,Plötzing2014}. Here, $<8$ femtosecond extreme-ultraviolet pulses are used to detect this imbalance, tracking both the number of excited electrons and their kinetic energy with time- and angle-resolved photoemission spectroscopy. Over a time window of approximately 25 fs after absorption of the pump pulse, we observe an increase in conduction band carrier density and a simultaneous decrease of the average carrier kinetic energy, revealing that relaxation is in fact dominated by inverse Auger scattering. Measurements of carrier scattering at extreme timescales by photoemission will serve as a guide to ultrafast control of electronic properties in solids for PetaHertz electronics \cite{Schultze2013}.}
\end{abstract}

\maketitle

\begin{figure}
	\center
  \includegraphics[width = 0.7\columnwidth]{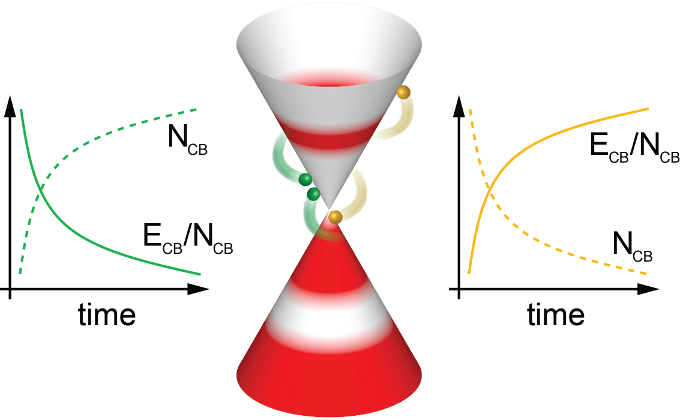}
  \caption{Different Coulomb-interaction-mediated scattering processes in photo-excited graphene: impact ionization (green) and Auger heating (yellow). These processes can be identified experimentally by comparing the total number of electrons inside the conduction band, $N_{CB}$, with the average kinetic energy of electrons inside the conduction band, $E_{CB}/N_{CB}$. Occupied states are shown in red, empty states are shown in white.}
  \label{fig1}
\end{figure}

The dynamics of pre-thermal Dirac carriers in graphene are expected to host interesting and unconventional phenomena. Microscopic simulations \cite{Winzer2010,Winzer2012} predict that different scattering events mediated by the Coulomb interaction may contribute to the ultrafast redistribution of the photo-excited electrons (see Fig. \ref{fig1}). For example, these theories imply that inverse Auger scattering (also known as impact ionization) may dominate at early times.

This can be understood as follows \cite{Winzer2010,Winzer2012}. As shown in Fig. \ref{fig1}, for a non-equilibrium distribution with holes in the valence band at $E = E_D - \frac{1}{2} \hbar\omega_{\text{pump}}$ and electrons in the conduction band at $E = E_D + \frac{1}{2} \hbar\omega_{\text{pump}}$ (where $E_D$ and $\hbar\omega_{\text{pump}}$ are the Dirac point and pump photon energy, respectively) in undoped graphene, the recombination of electron-hole pairs required for Auger heating is strongly suppressed due to the lack of holes at the top of the valence band. On the other hand, the available phase space for the inverse process (impact ionization), where the excess energy of an electron high in the conduction band is used to generate secondary electron-hole pairs, is large. Therefore, impact ionization is believed to dominate over Auger heating for as long as it takes to establish a thermalized electronic distribution. The resulting carrier multiplication, for which the absorption of a single photon may generate multiple electron-hole pairs, has raised interest for possible applications in photovoltaic devices.

However, these theoretical arguments apply only for undoped graphene and low excitation fluences (few $\mu$J/cm$^2$). At high pump fluences, the balance between impact ionization and Auger heating is predicted to be reestablished within a few tens of femtoseconds, reducing the carrier multiplication factor considerably \cite{Winzer2010,Winzer2012}. Furthermore, real graphene samples typically rest on a substrate resulting in a non-negligible doping of the graphene layer. The presence of either hole or electron doping will shift the relative importance of impact ionization and Auger heating, suppressing carrier multiplication.

Previous optical experiments deduced transient electron distribution functions by comparing differential transmission, reflectivity, or absorption data to model calculations. In this way, high-temporal-resolution experiments estimated an electronic thermalization time between 13 and 50 fs \cite{Breusing2011,Brida2013}. Furthermore, indirect evidence for carrier multiplication with $\sim$200 fs pulses in the low fluence regime ($<$ 30 $\mu$J/cm$^2$) was obtained by comparing the number of absorbed photons to the number of electron-hole pairs \cite{Plötzing2014} with some indications already in \cite{Brida2013}.

Photoemission techniques, which directly measure electron numbers as a function of energy and momentum, are ideally suited for a direct visualization of Auger scattering. Previous time- and angle-resolved photoemission spectroscopy (tr-ARPES) experiments were performed at pump fluences on the order of mJ/cm$^2$ and thus short time scales for the initial thermalization. The temporal resolution of $\geq$ 30 fs was then insufficient to resolve pre-thermal carrier distributions \cite{Johannsen2013,Gierz2013,Gierz2014,Ulstrup2014,Johannsen2015,Ulstrup2015,Gierz2015_1,Gierz2015_2}.
 
Here, we use tr-ARPES with $\leq$ 10 fs pump and probe pulses to access charge carrier dynamics in photo-excited graphene. The output of a 1 kHz, 30 fs Titanium:Sapphire amplifier operating at 1.55 eV photon energy was spectrally broadened in a neon-filled hollow-core fiber and recompressed using a set of chirped mirrors, resulting in 8 fs pulses with a slightly blue-shifted spectrum. 400 $\mu$J of energy were converted into extreme ultraviolet (XUV) pulses by High-order Harmonics Generation (HHG) in Argon. A single harmonic at $\hbar\omega_{\text{probe}}=30$ eV was selected with a time-preserving grating monochromator \cite{Frasetto2011} and used as a probe pulse for photoemission. The sample was excited with a fluence of 20 mJ/cm$^2$ with a pump pulse duration of $\sim$ 10 fs. Both pump and probe pulses were s-polarized, with the electric field vector in the plane of the graphene sample perpendicular to the $\Gamma$K direction. 

Tr-ARPES experiments were performed on lightly hole-doped quasi-freestanding epitaxial graphene monolayers with the Dirac point 200 meV above the equilibrium chemical potential \cite{Riedl2009}. These measurements were taken with two different monochromator gratings, optimizing either time or energy resolution \cite{Frasetto2011}.

\begin{figure}
	\center
  \includegraphics[width = 1\columnwidth]{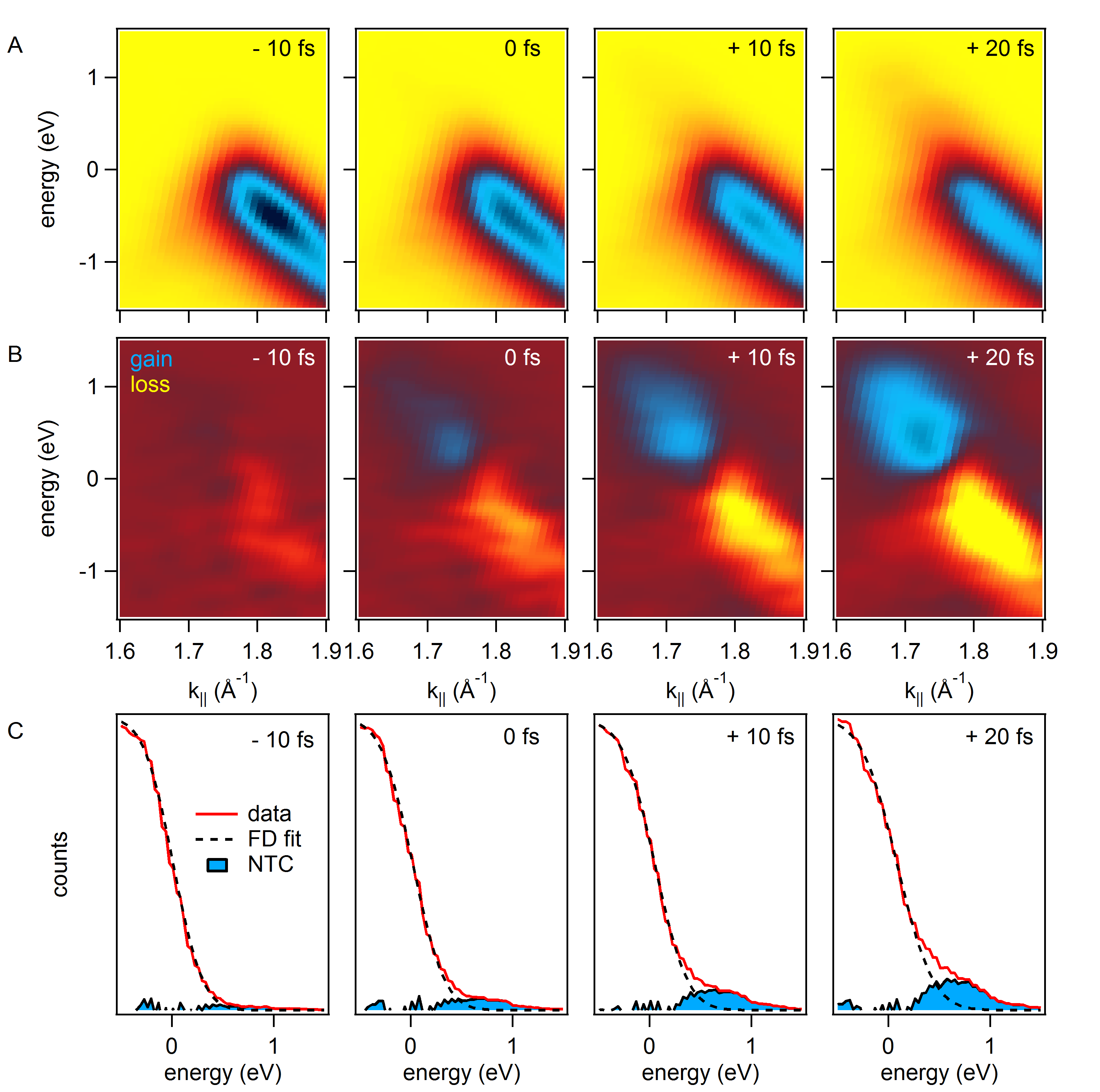}
  \caption{tr-ARPES measurements. A Snapshots for different pump-probe time delays. B Corresponding pump-induced changes of the photocurrent. The data in panel A and B has been smoothed. C Electronic distribution functions (red line) together with Fermi-Dirac fits (FD fit, black dashed line) revealing the presence of non-thermalized carriers (NTC, blue). The y-axis scale in C is linear. The measurements presented in this figure were carried out using the high-energy-resolution grating G300 (details see text).}
  \label{fig2}
\end{figure}

In Fig. \ref{fig2}A we show tr-ARPES snapshots of graphene's linear $\pi$-bands along the $\Gamma$K direction for selected pump-probe time delays across the rising edge of the pump-probe signal. Due to the s-polarization of the probe pulse \cite{Gierz2011} and to photoelectron interference effects \cite{Shirley1995} only the right-hand branch of the Dirac cone was visible in this geometry. These snapshots were recorded with a temporal resolution of 14 fs and an energy resolution of 500 meV (details see Supplementary Information) achieved by using a monochromator grating with 300 grooves per millimeter (G300) \cite{Frasetto2011}. The optical matrix element describing the absorption of the pump photon is anisotropic with nodes along the direction of the pump polarization (in this case perpendicular to the $\Gamma$K direction) and maxima in the direction perpendicular to the pump polarization (in this case $\Gamma$K) \cite{Malic2011,Mittendorff2014,Yan2014}. We probed the response of the electronic structure along the $\Gamma$K direction where the effect of the pump pulse is the strongest.
 
The pump-induced changes of the photocurrent are plotted in Fig. \ref{fig2}B. A loss (gain) of electrons below (above) the equilibrium chemical potential, which is used here as zero-energy reference, is observed. 

From the snapshots in Fig. \ref{fig2}A, transient electron distribution functions were extracted and compared to Fermi-Dirac (FD) distributions in Fig. \ref{fig2}C. At negative delays, immediately before arrival of the pump pulse (-10 fs in Fig. \ref{fig2}C), the distribution follows a FD distribution, indicating the presence of a completely thermalized electron gas. Near zero time delay, a shoulder develops in the conduction band above $E_D=200$ meV. This shoulder cannot be fitted with a FD distribution. The residual weight between the experimental distribution function (red line in Fig. \ref{fig2}C) and the FD fit (dashed black line in Fig. \ref{fig2}C) is taken as proportional to the number of non-thermal carriers (NTCs, blue shaded area in Fig. \ref{fig2}C). We find that the number of NTCs keeps increasing until the peak of the pump-probe signal is reached at a time delay of $\sim$20 fs. These NTCs can be understood as a precursor of the inverted carrier population observed in previous studies with longer pump pulses at slightly smaller photon energies \cite{Li2012,Gierz2013,Gierz2015_1}.

\begin{figure}
	\center
  \includegraphics[width = 0.5\columnwidth]{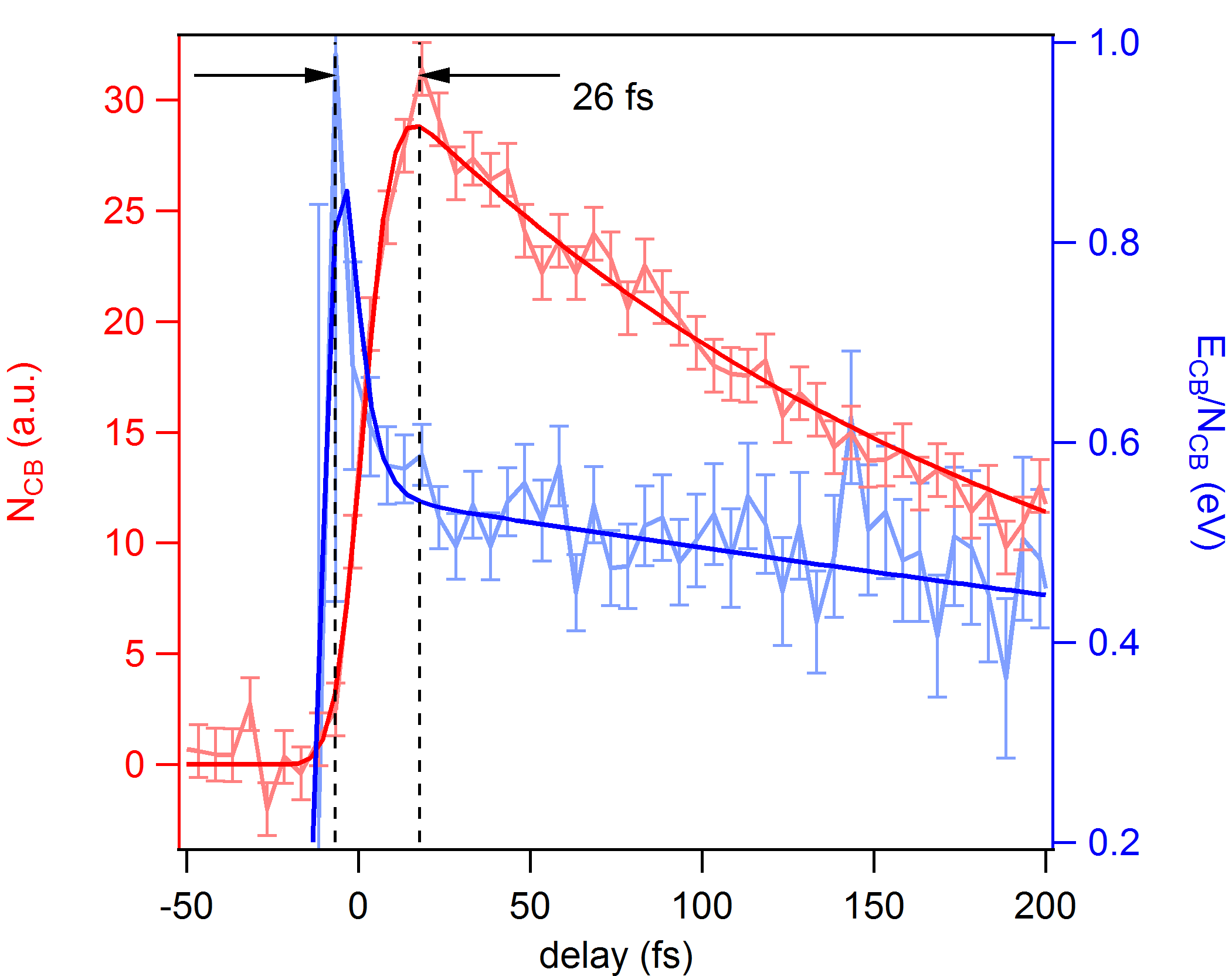}
  \caption{Direct evidence for impact ionization. Comparison between the temporal evolution of the total number of carriers inside the conduction band ($N_{CB}$, light red) and the temporal evolution of their average kinetic energy ($E_{CB}/N_{CB}$, light blue). Around zero time delay $E_{CB}/N_{CB}$ already decreases while $N_{CB}$ keeps increasing, indicating impact ionization. Dark red and blue lines are fits to the data that serve as guides to the eye. The fitting function consists of an error function to describe the rising edge plus a single ($N_{CB}$) or double exponential decay ($E_{CB}/N_{CB}$). The measurements presented in this figure were carried out using the low-energy-resolution grating G60 (details see text). Error bars represent the standard deviation.}
  \label{fig3}
\end{figure}

Higher-temporal-resolution measurements were performed with a second monochromator grating with 60 grooves per millimeter (G60), delivering in a temporal resolution of 8 fs and an energy resolution of 800 meV (details see Supplementary Information). The number of carriers in the conduction band at $E > E_D$, $N_{CB}$, and their average kinetic energy, $E_{CB}/N_{CB}$, were determined directly from the raw data (details see Methods and Supplementary Information) and displayed in Fig. \ref{fig3}. For about 25 fs around zero pump-probe time delay, the average kinetic energy $E_{CB}/N_{CB}$ was observed to decrease while the number of carriers $N_{CB}$ kept increasing, indicating impact ionization. 

From these measurements the following scenario can be envisaged. Impact ionization is the primary scattering mechanism during the first $\sim$25 fs, accumulating carriers at the bottom of the conduction band and establishing a precursor of the population inversion observed previously \cite{Li2012,Gierz2013,Gierz2015_1}. This state then likely decays through Auger heating and electron-phonon scattering within $\sim$100 fs, re-establishing a single Fermi Dirac distribution \cite{Gierz2013}. We note that the temporal evolution of the electronic temperature and the decay time of the non-thermal carriers found in our experiment (see Supplementary Information) further substantiate this interpretation.

In summary, we have used tr-ARPES with $\leq$ 10 fs pulses to identify the primary scattering events that result in the rapid thermalization of photo-excited electron-hole pairs in graphene. By comparing the number of electrons inside the conduction band with their average kinetic energy we find that impact ionization is the predominant scattering channel within the first $\sim$25 fs also in doped graphene at high fluence. Whether the observed carrier multiplication can be exploited for solar cell applications remains questionable, as the absence of a band gap in graphene makes charge separation difficult. Nevertheless, the ultrafast dynamics of photo-excited electron-hole pairs observed here will be of use for the design of new electronic and optoelectronic devices operating at PHz rates \cite{Schultze2013} based on graphene and other materials.

\section{Methods:}

{\bf Samples.} Quasi-freestanding epitaxial graphene samples on silicon carbide were grown as previously described in \cite{Riedl2009}. Prior to graphene growth the silicon carbide substrates were etched in hydrogen atmosphere to remove scratches from mechanical polishing. In a second step the substrates were annealed in argon atmosphere resulting in the growth of one carbon monolayer on the silicon-terminated face of the substrate. This carbon monolayer was subsequently decoupled from the substrate by hydrogen intercalation and characterized by static ARPES measurements. The samples were transported to the Artemis user facility under ambient conditions, reinserted into ultra-high vacuum, and cleaned by a mild annealing, recovering the original band structure.

{\bf Tr-ARPES.} Tr-ARPES experiments were performed using the Materials Science end station at the Artemis user facility at the Rutherford Appleton Laboratory in Harwell, United Kingdom. The setup consists of a Titanium:Sapphire amplifier ($\hbar\omega=1.55$\,eV) operating at 1 kHz with 30 fs pulse duration. 1 mJ of energy was compressed down to 8 fs with a hollow-core fiber filled with 1.4 bar of neon operated with a pressure gradient and followed by 10 chirped mirrors. Pulse compression resulted in a slight blue shift of the spectrum. 400 $\mu$J of energy were used for high harmonics generation (HHG) in argon, 20 mJ/cm$^2$ were used to excite the sample. Due to the longer path in air two additional chirped mirrors were added to the pump path resulting in a pump pulse duration of 10 fs. All pulse durations were measured using second-harmonic frequency-resolved optical gating (FROG). Out of the broad HHG spectrum one particular energy, $\hbar\omega_{\text{probe}}=30$ eV, is selected by a time-preserving grating monochromator \cite{Frasetto2011} and used as a probe pulse for photoemission. For the present investigation two different gratings with different dispersion (G300 with 300 grooves per mm and G60 with 60 grooves per mm) have been used for wavelength selection \cite{Frasetto2011}.

{\bf Data analysis.} The data in Fig. \ref{fig2}A and B has been smoothed. To obtain the electron distribution functions in Fig. \ref{fig2}C we took lineouts at constant angle (energy distribution curves) from the raw data, determined their integrated intensity within the full width at half maximum (FWHM) of the peak and attributed the resulting number to the energy of the peak position. The number of carriers in the conduction band in Fig. 3 was obtained by integrating the electron distributions for $E>E_D$. The kinetic energy of carriers in the conduction band in Fig. \ref{fig3} was obtained by integrating the electron distributions multiplied by the energy measured with respect to the Dirac point for $E>E_D$.

\section{Acknowledgments:}

We thank A. S. Wyatt, O. Alexander, and P. Rice for technical support during the beamtime, and J. Harms for support with the figures. This work was supported by the German Research Foundation (DFG) in the framework of the Priority Program 1459 `Graphene'. Access to the Artemis facility at the Rutherford Appleton Laboratory was funded by STFC.
 
\section{Author contributions:}

I.G. conceived the research project. I.G., F.C., S.A., M.C.C., and C.C. performed the tr-ARPES experiments. S.L. has grown the samples, S.L. and S.A. characterized the samples with static ARPES. S.A., I.G., and C.R.A. anaylized the data. I.G., F.C., S.A., and A.C. interpreted the data. I.G. and A.C. wrote the manuscript with input from all coauthors.

\section{Competing financial interests:}

The authors declare that they have no competing financial interests.

\pagebreak

\section{Supplementary Information:}

\subsection{Sample characterization}

\begin{figure}[h]
	\center
  \includegraphics[width = 0.5\columnwidth]{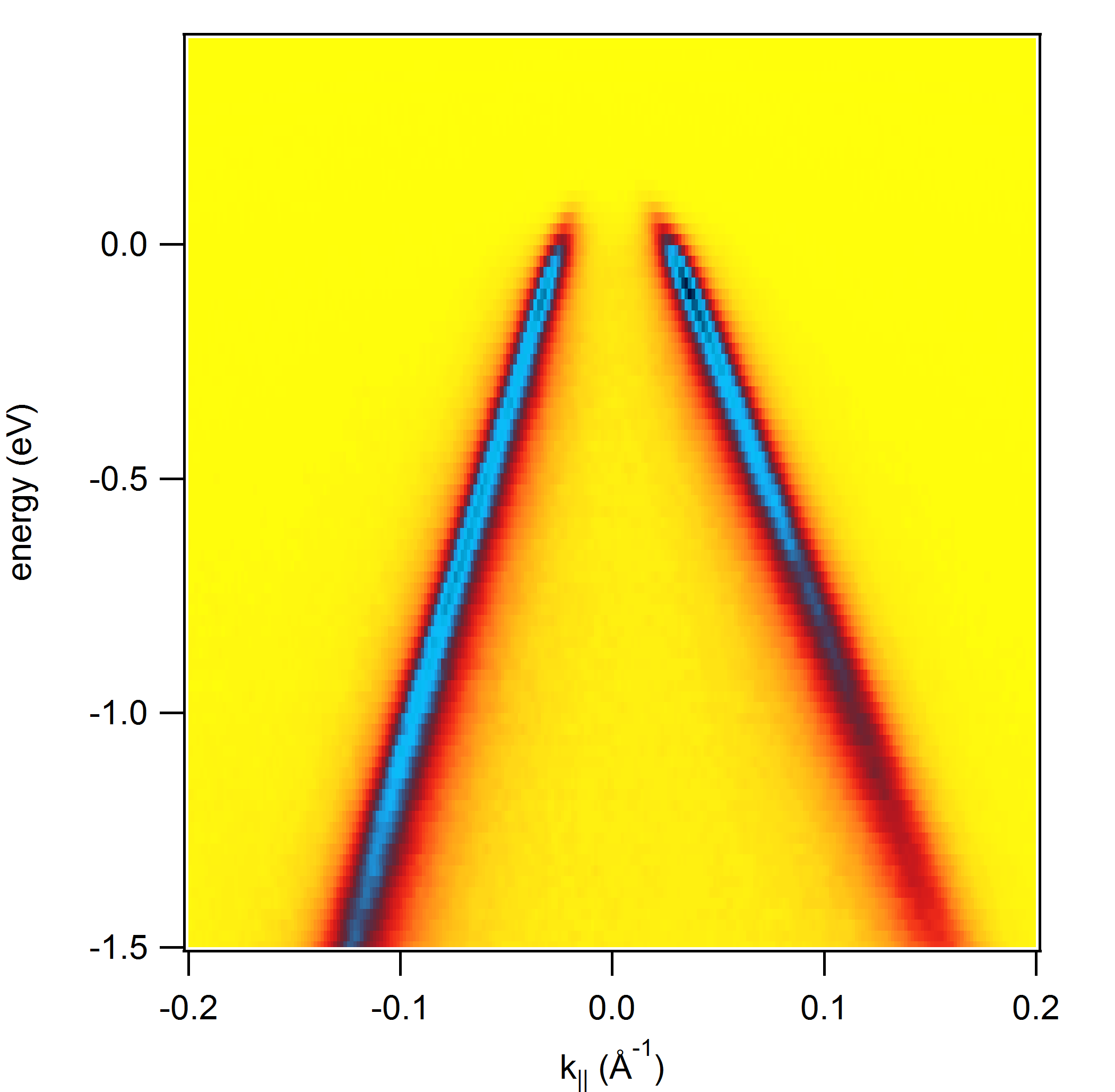}
  \caption{High-resolution ARPES spectrum of hydrogen-intercalated monolayer graphene on silicon carbide measured along a cut through the K point perpendicular to the $\Gamma$K direction.}
  \label{figS1}
\end{figure}

After growth and hydrogen intercalation the graphene samples were characterized using high-resolution angle-resolved photoemission spectroscopy (ARPES). In Fig. \ref{figS1} we show the electronic structure along a cut through the K point perpendicular to the $\Gamma$K direction. The sample is lightly hole-doped with the Dirac point $\sim$200 meV above the equilibrium chemical potential.

\subsection{Pulse compression}

\begin{figure}
	\center
  \includegraphics[width = 1\columnwidth]{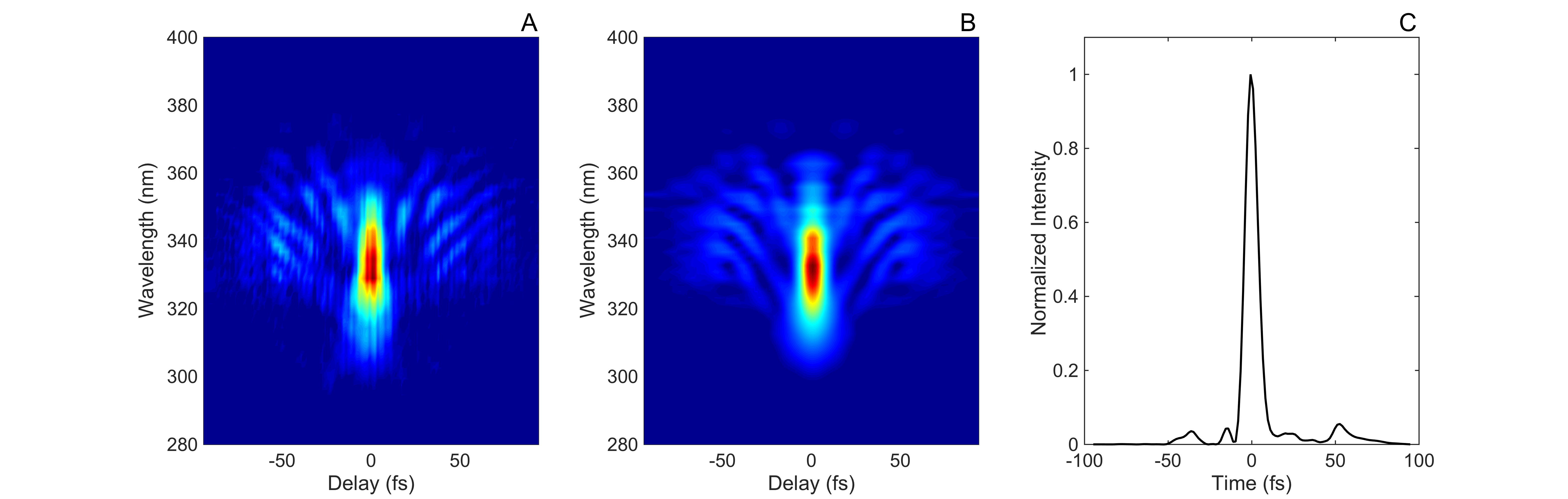}
  \caption{Second-harmonic FROG reconstruction of the compressed pulses. A Second-harmonic FROG trace acquired as a function of delay. B Reconstructed second-harmonic FROG trace. C Reconstructed pulse intensity profile. The retrieved pulse duration is 8.3 fs.}
  \label{figS2}
\end{figure}

Pulse compression from 30 fs down to 8 fs was achieved by broadening the output spectrum of a Titanium:Sapphire amplifier in a 1.4 bar neon-filled hollow-core fiber operated with a pressure gradient and subsequent recompression using 10 chirped mirrors. The pulse duration was measured with second-harmonic frequency-resolved optical gating (FROG). For that purpose the pulse was split by a 50/50 beam splitter, and the two beams were overlapped again on a BBO crystal. The generated second harmonic signal was analyzed with a spectrometer. Figure \ref{figS2}A shows the two-dimensional FROG trace acquired as a function of delay between the two pulses. The reconstructed FROG trace and pulse intensity profile are reported in panels B and C of the same figure, respectively. The retrieved pulse duration is 8.3 fs. The delay-integrated spectrum is centered around 335 nm (3.70 eV), indicating a central wavelength of the compressed pulse of 670 nm (1.85 eV).

This beam was split with an 80/20 beam splitter and used for high harmonics generation (HHG) in argon (400 $\mu$J) and to excite the sample (3.5 $\mu$J), respectively. Due to the longer path in air, two more chirped mirrors were added to the pump path, resulting in a pump pulse duration of 10 fs.

An estimate for the temporal resolution of the tr-ARPES experiment can be obtained from the rise time of the pump-probe signal. For that purpose, the rising edge of the number of non-thermal carriers (see below) has been fitted with an error function, $\text{erf}((t-t_0)/\text{rt})$. The rise time, rt, is 14 $\pm$ 2 fs for grating G300 (300 grooves per mm) and 8 $\pm$ 1 fs for grating G60 (60 grooves per mm). This rise time is related to the full width at half maximum (FWHM) of the cross correlation between pump and probe pulses via $\text{FWHM}=1.665\times\text{rt}$. The energy resolution has been obtained by fitting electron distribution functions (see below) at negative delay with a Fermi-Dirac distribution convoluted by a Gaussian. The temperature was fixed to 300 K (the temperature of the sample). The FWHM of the Gaussian was obtained as a fitting parameter, giving FWHM = 500 meV for G300 and FWHM = 800 meV for G60.

\subsection{Extraction of electronic distribution functions}

The energy resolution of the tr-ARPES measurements was restricted to 500 meV using grating G300 in Fig. \ref{fig2} and 800 meV using grating G60 in Fig. 3 of the main manuscript. This means that the established standard methods to extract electron distribution functions (simple momentum integration \cite{Gierz2013} or MDC method described in \cite{Ulstrup2014_2}) fail, meaning that the equilibrium distribution function determined that way does not follow a Fermi-Dirac distribution.
 
\begin{figure}
	\center
  \includegraphics[width = 1\columnwidth]{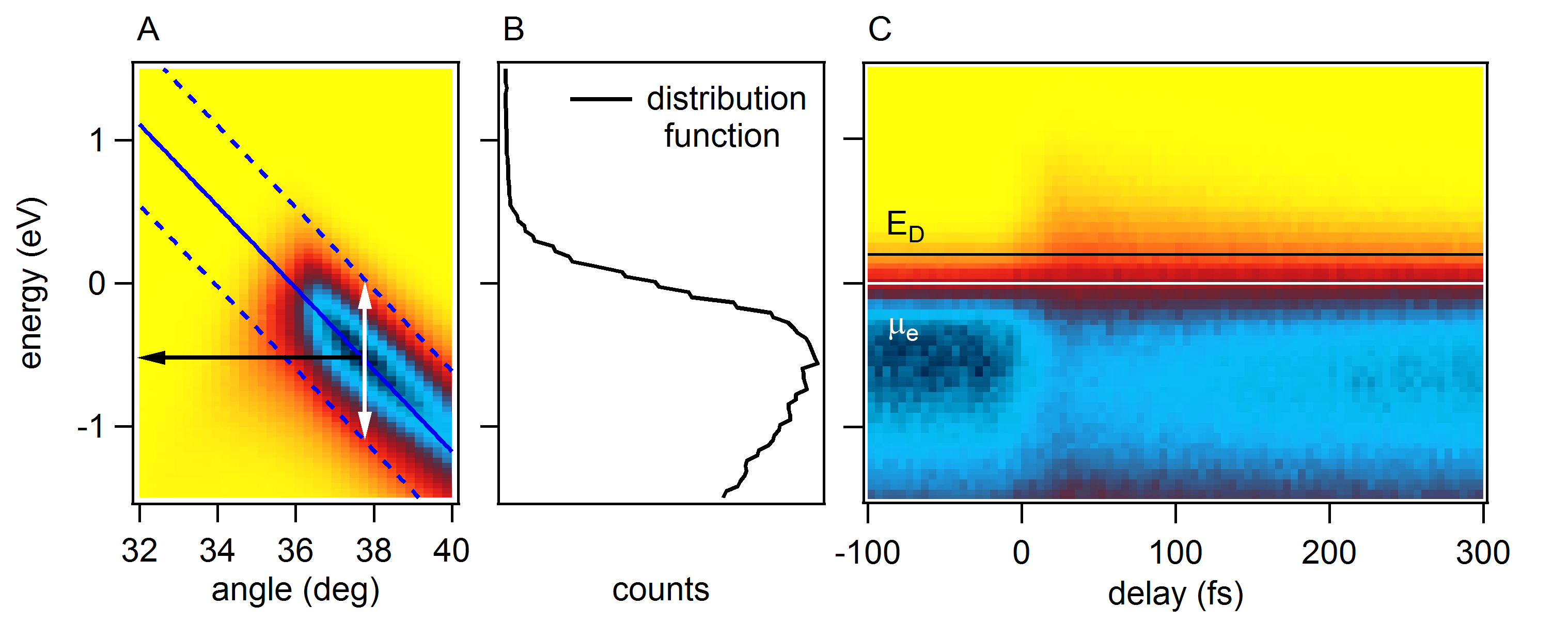}
  \caption{A Electron distributions in B were extracted from the raw data by summing up the intensity along the white arrow and attributing it to the energy indicated by the black arrow. Blue lines indicate the peak position (continuous) $\pm$ full width at half maximum (dashed). B Electron distribution function as a function of energy. The drop at negative energy is caused by photoemission matrix element effects. C Electron distribution functions from B for different pump-probe delays as a two-dimensional color plot. The position of the chemical potential $\mu_e$ and the Dirac point $E_D$ are indicated by white and black lines, respectively.}
  \label{figS3}
\end{figure}

Our method is based on the assumption that the comparatively low energy resolution in our experiment broadens the data along the energy axis, while the angular resolution is unaffected. Thus, we took lineouts at constant angle (energy distribution curves, EDCs), determined their integrated intensity within the FWHM of the peak and attributed the resulting number to the energy of the peak position. This procedure is illustrated in Fig. \ref{figS3}. The resulting distribution functions are shown in Fig. \ref{fig2}C of the main manuscript.

From similar distribution functions recorded using G60 we calculated the number of carriers inside the conduction band, $N_{CB}$, and their average kinetic energy, $E_{CB}/N_{CB}$, as described in the following. In Fig. \ref{figS3}C the electronic distribution is plotted as a function of energy and pump-probe delay. Due to the light hole-doping the bottom of the conduction band, $E_D$, lies $\sim$200 meV above the equilibrium chemical potential, $\mu_e$. $N_{CB}$ is determined by summing up the intensity between $E = E_D$ and $E = 2.6$ eV for a given delay. Similarly, $E_{CB}$ is determined by summing up the intensity multiplied by the energy referenced with respect to the Dirac point ($E_D = 0$ eV) between $E = E_D$ and $E = 2.6$ eV for a given delay.

\subsection{Results of Fermi-Dirac fits}

\begin{figure}
	\center
  \includegraphics[width = 0.5\columnwidth]{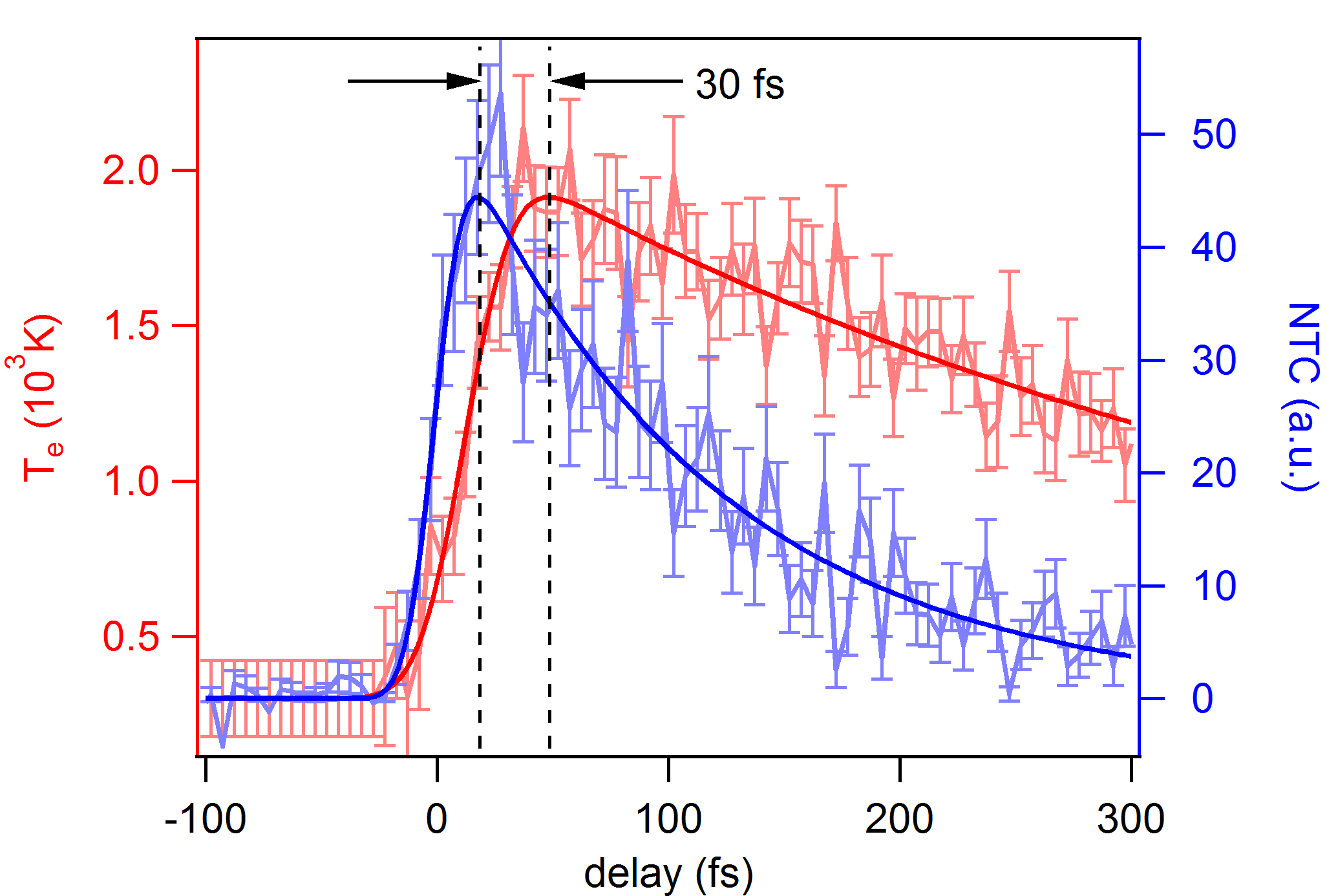}
  \caption{Results of Fermi-Dirac fit of the data shown in Fig. \ref{fig2}C of the main article. Electronic temperature ($T_e$, light red) and number of non-thermalized carriers (NTC, light blue) as a function of pump-probe delay. Zero time delay is defined as the middle of the rising edge of the number of non-thermalized carriers. Dark red and blue lines are fits to the data. Error bars represent the standard deviation.}
  \label{figS4}
\end{figure}

Figure \ref{figS4} summarizes the results of the Fermi-Dirac fits from Fig. \ref{fig2}C in the main manuscript. The figure shows the temporal evolution of the electronic temperature ($T_e$, light red), and of the number of non-thermal carriers (NTC, light blue), where NTC corresponds to the area of the blue shaded region in Fig. \ref{fig2}C of the main manuscript. Dark red and blue lines represent fits to the data including an error function to describe the rising edge and a single exponential decay.

The temperature peaks 30 fs after the number of non-thermal carriers reaches its maximum. The decay times of the temperature and of the number of non-thermal carriers are 412 $\pm$ 32 fs and 113 $\pm$ 6 fs, respectively. In graphene, the elevated electronic temperature is known to cool down by optical and acoustic phonon emission \cite{Yan2009,Kang2010,Song2012}. The life time of the non-thermal carriers is similar to the one found for the population inversion in \cite{Li2012,Gierz2013,Gierz2015_1}, indicating that complete thermalization is hindered by the relaxation bottleneck imposed by the Dirac point.

\subsection{tr-ARPES snapshots for grating G60}

\begin{figure}
	\center
  \includegraphics[width = 1\columnwidth]{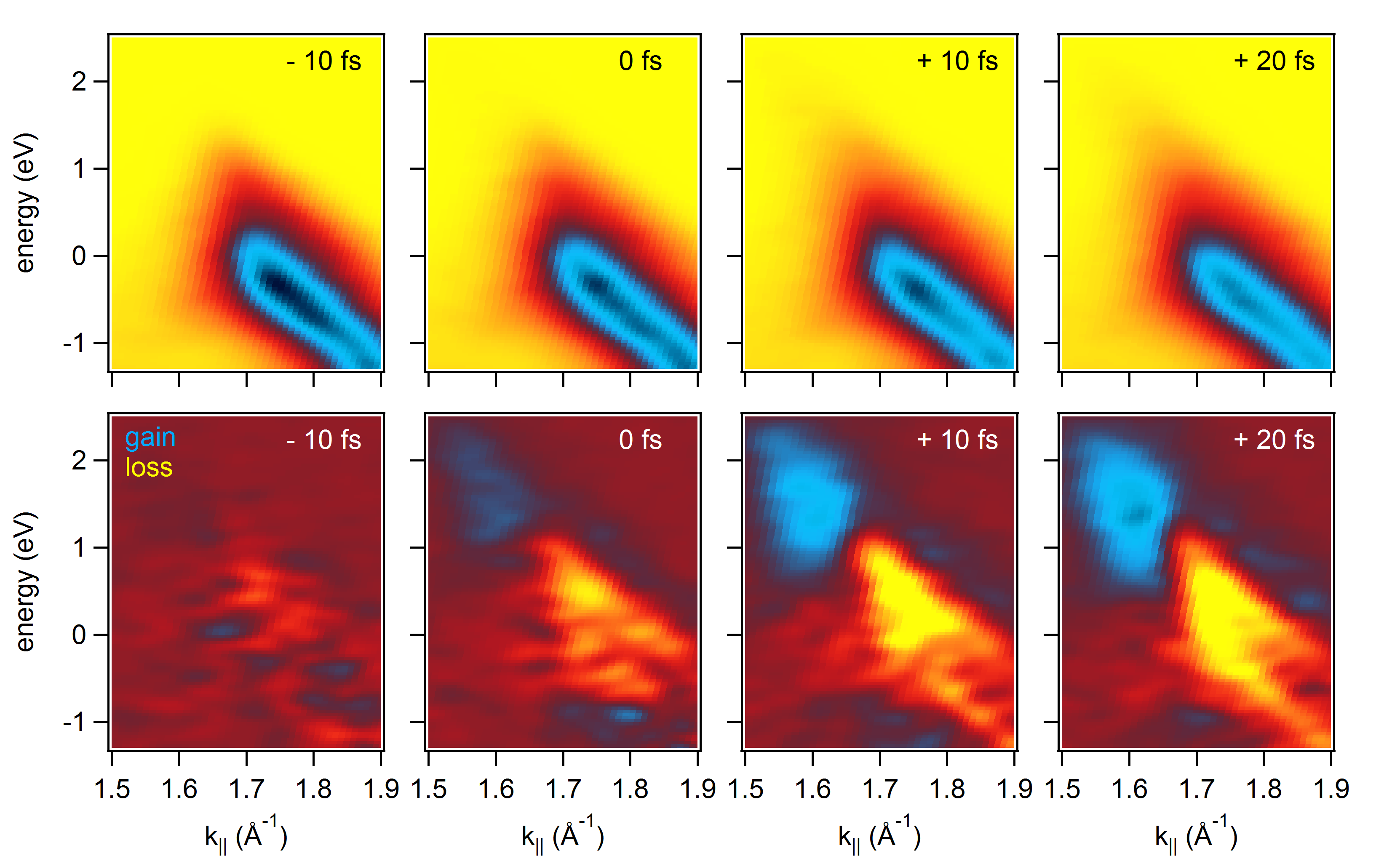}
  \caption{tr-ARPES snapshots recorded with grating G60.}
  \label{figS5}
\end{figure}

The tr-ARPES data recorded with grating G60 that has been used to generate Fig. \ref{fig3} in the main manuscript is displayed in Fig. \ref{figS5}. Similar to the data presented in Fig. \ref{fig2}A and B the data in Fig. \ref{fig3} has been smoothed. Data displayed in Fig. \ref{fig2}C and Fig. \ref{fig3} in the main manuscript and Fig. \ref{figS4} of the Supplementary Information was extracted directly from the un-smoothed raw data as explained above.

\begin{thebibliography}{12}

\bibitem{Winzer2010} Winzer, T., Knorr, A. \& Malic, E. Carrier Multiplication in Graphene. {\it Nano Lett.} {\bf 10}, 4839 (2010).
\bibitem{Winzer2012} Winzer, T. \& Malic, E. Impact of Auger processes on carrier dynamics in graphene. {\it Phys. Rev. B} {\bf 85}, 241404(R) (2012).
\bibitem{Brida2013} Brida, D., Tomadin, A., Manzoni, C., Kim, Y. J., Lombardo, A., Milana, S., Nair, R. R., Novoselov, K. S., Ferrari, A. C., Cerullo, G. \& Polini, M. Ultrafast collinear scattering and carrier multiplication in graphene. {\it Nat. Commun.}, DOI: 10.1038/ncomms2987 (2013).
\bibitem{Plötzing2014} Pl{\"o}tzing, T., Winzer, T., Malic, E., Neumaier, D., Knorr, A. \& Kurz, H. Experimental Verification of Carrier Multiplication in Graphene. {\it Nano Lett.} {\bf 14}, 5371 (2014).
\bibitem{Schultze2013} Schultze, M., Bothschafter, E. M., Sommer, A., Holzner, S., Schweinberger, W., Fiess, M., Hofstetter, M., Kienberger, R., Apalkov, V., Yakovlev, V. S., Stockman, M. I. \& Krausz, F. Controlling dielectrics with the electric field of light. {\it Nature} {\bf 493}, 75 (2013).
\bibitem{Breusing2011} Breusing, M., Kuehn, S., Winzer, T., Malic, E., Milde, F., Severin, N., Rabe, J. P., Ropers, C., Knorr, A. \& Elsaesser, T. Ultrafast nonequilibrium carrier dynamics in a single graphene layer. {\it Phys. Rev. B} {\bf 83}, 153410 (2011).
\bibitem{Johannsen2013} Johannsen, J. C., Ulstrup, S., Cilento, F., Crepaldi, A., Zacchigna, M., Cacho, C., Turcu, I. C. E., Springate, E., Fromm, F., Raidel, C., Seyller, T., Parmigiani, F., Grioni, M. \& Hofmann, P. Direct View of Hot Carrier Dynamics in Graphene. {\it Phys. Rev. Lett.} {\bf 111}, 027403 (2013).
\bibitem{Gierz2013} Gierz, I., Petersen, J. C., Mitrano, M., Cacho, C., Turcu, I. C. E., Springate, E., St\"ohr, A., K\"ohler, A., Starke, U. \& Cavalleri, A. Snapshots of non-equilibrium Dirac carrier distributions in graphene. {\it Nat. Mater.} {\bf 12}, 1119 (2013).
\bibitem{Gierz2014} Gierz, I., Link, S., Starke, U. \& Cavalleri, A. Non-equilibrium Dirac carrier dynamics in graphene investigated with time- and angle-resolved photoemission spectroscopy. {\it Faraday Disc.} {\bf 171}, 311 (2014).
\bibitem{Ulstrup2014} Ulstrup, S., Johannsen, J. C., Cilento, F., Miwa, J. A., Crepaldi, A., Zacchigna, M., Cacho, C., Chapman, R., Springate, E., Mammadov, S., Fromm, F., Raidel, C., Seyller, T., Parmigiani, F., Grioni, M., King, P. D. C. \& Hofmann, P. Ultrafast Dynamics of Massive Dirac Fermions in Bilayer Graphene. {\it Phys. Rev. Lett.} {\bf 112}, 257401 (2014).
\bibitem{Johannsen2015} Johannsen, J. C., Ulstrup, S., Crepaldi, A., Cilento, F., Zacchigna, M., Miwa, J. A., Cacho, C., Chapman, R. T., Springate, E., Fromm, F., Raidel, C., Seyller, T., King, P. D. C., Parmigiani, F., Grioni, M. \& Hofmann, P. Tunable Carrier Multiplication and Cooling in Graphene. {\it Nano Lett.} {\bf 15}, 326 (2015).
\bibitem{Ulstrup2015} Ulstrup, S., Johannsen, J. C., Crepaldi, A., Cilento, F., Zacchigna, M., Cacho, C., Chapman, R. T., Springate, E., Fromm, F., Raidel, C., Seyller, T., Parmigiani, F., Grioni, M. \& Hofmann, P. Ultrafast electron dynamics in epitaxial graphene investigated with time- and angle-resolved photoemission spectroscopy. {\it J. Phys.: Condens. Matter} {\bf 27}, 164206 (2015).

\bibitem{Gierz2015_1} Gierz, I., Mitrano, M., Petersen, J. C., Cacho, C., Turcu, I. C. E., Springate, E., St\"ohr, A., K\"ohler, A., Starke, U. \& Cavalleri, A. Population inversion in monolayer and bilayer graphene. {\it J. Phys.: Condens. Matter} {\bf 27}, 164204 (2015).
\bibitem{Gierz2015_2} Gierz, I., Mitrano, M., Bromberger, H., Cacho, C., Chapman, R., Springate, E., Link, S., Starke, U., Sachs, B., Eckstein, M., Wehling, T. O., Katsnelson, M. I., Lichtenstein, A. \& Cavalleri, A. Phonon-Pump Extreme-Ultraviolet-Photoemission Probe in Graphene: Anomalous Heating of Dirac Carriers by Lattice Deformation. {\it Phys. Rev. Lett.} {\bf 114}, 125503 (2015).

\bibitem{Frasetto2011} Frassetto, F., Cacho, C., Froud, C. A., Turcu, I. C. E., Villoresi, P., Bryan, W. A., Springate, E. \& Poletto, L. Single-grating monochromator for extreme-ultraviolet ultrashort pulses. {\it Opt. Express} {\bf 19}, 19169 (2011).
\bibitem{Riedl2009} Riedl, C., Coletti, C., Iwasaki, T., Zakharov, A. A. \& Starke, U. Quasi-Free-Standing Epitaxial Graphene on SiC Obtained by Hydrogen Intercalation. {\it Phys. Rev. Lett.} {\bf 103}, 246804 (2009).
\bibitem{Gierz2011} Gierz, I., Henk, J., H\"ochst, H., Ast, C. R. \& Kern, K. Illuminating the dark corridor in graphene: Polarization dependence of angle-resolved photoemission spectroscopy on graphene. {\it Phys. Rev. B} {\bf 83}, 121408(R) (2011).
\bibitem{Shirley1995} Shirley, E. L., Terminello, L. J., Santoni, A. \& Himpsel, F. J. Brillouin-zone-selection effects in graphite photoelectron angular distributions. {\it Phys. Rev. B} {\bf 51}, 13614 (1995).
\bibitem{Malic2011} Malic, E., Winzer, T., Bobkin, E. \& Knorr, A. Microscopic theory of absorption and ultrafast many-particle kinetics in graphene. {\it Phys. Rev. B} {\bf 84}, 205406 (2011).
\bibitem{Mittendorff2014} Mittendorff, M., Winzer, T., Malic, E., Knorr, A., Berger, C., de Heer, W. A., Schneider, H., Helm, M. \& Winnerl, S. Anisotropy of Excitation and Relaxation of Photogenerated Charge Carriers in Graphene. {\it Nano Lett.} {\bf 14}, 1504 (2014).
\bibitem{Yan2014} Yan, X.-Q., Yao, J., Liu, Z.-B., Zhao, X., Chen, X.-D., Gao, C., Xin, W., Chen, Y. \& Tian, J.-G. Evolution of anisotropic-to-isotropic photoexcited carrier distribution in graphene. {\it Phys. Rev. B} {\bf 90}, 134308 (2014).
\bibitem{Li2012} Li, T., Luo, L., Hupalo, M., Zhang, J., Tringides, M. C., Schmalian, J. \& Wang, J. Femtosecond Population Inversion and Stimulated Emission of Dense Dirac Fermions in Graphene. {\it Phys. Rev. Lett.} {\bf 108}, 167401 (2012).

\bibitem{Ulstrup2014_2} Ulstrup, S., Johannsen, J. C., Grioni, M. \& Hofmann, P. Extracting the temperature of hot carriers in time- and angle-resolved photoemission. {\it Rev. Sci. Instrum.} {\bf 85}, 013907 (2014).
\bibitem{Yan2009} Yan, H., Song, D., Malik, K. F., Chatzakis, I., Maultzsch, J. \& Heinz, T. F. Time-resolved Raman spectroscopy of optical phonons in graphite: Phonon anharmonic coupling and anomalous stiffening. {\it Phys. Rev. B} {\bf 80}, 121403(R) (2009).
\bibitem{Kang2010} Kang, K., Abdula, D., Cahill, D. G. \& Shim, M. Lifetimes of optical phonons in graphene and graphite by time-resolved incoherent anti-Stokes Raman scattering. {\it Phys. Rev. B} {\bf 81}, 165405 (2010).
\bibitem{Song2012} Song, J. C. W., Reizer, M. Y. \& Levitov, L. S. Disorder-Assisted Electron-Phonon Scattering and Cooling Pathways in Graphene. {\it Phys. Rev. Lett.} {\bf 109}, 106602 (2012).

\end{thebibliography}
\end{document}